# Automatic Post-Stroke Lesion Segmentation on MR Images using 3D Residual Convolutional Neural Network

Naofumi Tomita, MS[1], Steven Jiang, BS[2], Matthew E. Maeder, MD[3], Saeed Hassanpour, PhD[1,2,4]

[1]Department of Biomedical Data Science, Geisel School of Medicine at Dartmouth, Hanover, NH 03755, USA

[2]Department of Computer Science, Dartmouth College, Hanover, NH 03755, USA

[3]Department of Radiology, Dartmouth-Hitchcock Medical Center, Lebanon, NH 03756, USA

[4]Department of Epidemiology, Geisel School of Medicine at Dartmouth, Hanover, NH 03755, USA

## Abstract

In this paper, we demonstrate the feasibility and performance of deep residual neural networks for volumetric segmentation of irreversibly damaged brain tissue lesions on T1-weighted MRI scans for chronic stroke patients. A total of 239 T1-weighted MRI scans of chronic ischemic stroke patients from a public dataset were retrospectively analyzed by 3D deep convolutional segmentation models with residual learning, using a novel zoom-in&out strategy. Dice similarity coefficient (DSC), Average symmetric surface distance (ASSD), and Hausdorff distance (HD) of the identified lesions were measured by using the manual tracing of lesions as the reference standard. Bootstrapping was employed for all metrics to estimate 95% confidence intervals. The models were assessed on the test set of 31 scans. The average DSC was 0.64 (0.51-0.76) with a median of 0.78. ASSD and HD were 3.6 mm (1.7-6.2 mm) and 20.4 mm (10.0-33.3 mm), respectively. To the best of our knowledge, this performance is the highest achieved on this public dataset. The latest deep learning architecture and techniques were applied for 3D segmentation on MRI scans and demonstrated to be effective for volumetric segmentation of chronic ischemic stroke lesions.

# Introduction

Stroke is one of the leading causes of long-term adult disability worldwide (1). Recent studies show that 36% to 71% of post-stroke survivors had a disability after at least five years (2-5). Rehabilitation is crucial for long-term functional recovery. The effectiveness of rehabilitation varies, however, because functional and structural changes in the brain differ among patients. Identifying the damaged brain network in patients would help clinicians to predict functional outcomes in response to targeted rehabilitation, which benefits patients by optimizing treatment resources and providing personal and efficient care (6-8). T1-weighted (T1W) magnetic resonance imaging (MRI) is the most common resource in research for chronic stroke lesions because lesions are visible on T1W images after a month and the produced images have high resolution. Tracing these lesions manually, however, is time-intensive and prone to errors (9).

Many approaches have been proposed for automatic segmentation of chronic lesions on T1W MRIs after a stroke (10-18). Compared to research on automatic segmentation of acute stroke lesions, however, methods for chronic lesion segmentation are under-explored. One major difference between the acute and chronic imaging segmentation is that the former utilizes diffusion—and/or perfusion—weighted imaging, while the latter typically uses high-resolution T1W imaging. Methods developed for acute stroke lesion segmentation are not readily applicable to chronic stroke analysis due to the different characteristics of these MIR pulse sequences and the high-resolution data of T1W MRIs.

Recently, convolutional neural networks (CNNs) have achieved expert-level performance in various radiology image analysis tasks (19-22). Three-dimensional (3D) CNNs are deep learning architectures that can extract 3D spatial features. Since diagnosing stroke lesions by neuroradiologists requires analysis of a lesion and its surrounding area (23), 3D CNNs are suitable for this task. This is because 3D CNNs incorporate the contextual information of voxels (i.e., volumetric pixels) into analysis by capturing both low-level local features (i.e., edges and corners) and high-level global features (i.e., the anatomy of brains).

In this study, we developed an effective deep learning model for 3D segmentation to identify areas of infarcted brain tissue on MRI images. To develop the method, we utilized a public dataset of chronic stroke lesions from patients' T1W MRI scans.

## Materials and Methods

**Data Source**

To develop and evaluate our algorithm in this study, we used a publicly available dataset of volumetric MRI scans of patient brains with anatomical tracings of lesions after stroke (ATLAS) (24). In the ATLAS dataset, a total of 304 MRI scans were collected. Stroke lesions on T1-weighted MRI images are manually traced and established by trained students and research fellows under the supervision of an expert tracer and a neuroradiologist. The collection of the ATLAS dataset and the subsequent sharing of the data were approved by the study's institutional review board (IRB). Informed consent was obtained from all subjects before the data collection. The ATLAS contains two subsets of data: 239 scans normalized to MNI-152 space and 229 scans normalized and further defaced. We use the 239 normalized non-defaced scans because of possible data variability and generalizability concerns induced by the defacing method applied to scans (24). The size of scans is $197 \times 233 \times 189$ mm$^3$ and the canonical size of a voxel is 1 mm$^3$. Lesion size in the dataset ranges from 10 mm$^3$ to $2.8 \times 10^5$ mm$^3$. Demographic data of the dataset is not available. The statistics of the dataset are summarized in Table E1 and E2 in the Appendix.

**3D Segmentation using Deep Convolutional Neural Network**

For the 3D brain lesion segmentation task, we use a 3D U-Net (25), which is the state-of-the-art deep learning architecture for volumetric segmentation tasks. U-Net architecture has characteristic internal skipping connections between layers to propagate information from earlier layers (encoder) to later layers (decoder). Figure 1 shows the overview of our 3D U-Net model in this study. We extended the 3D U-Net architecture to accommodate to our task and detailed the modification in the supplemental materials, Appendix E1. Our objective function $L$ is an affine combination of binary cross entropy (BCE) loss function and Dice loss function (26), which we describe in detail in the supplemental material, Appendix E2.

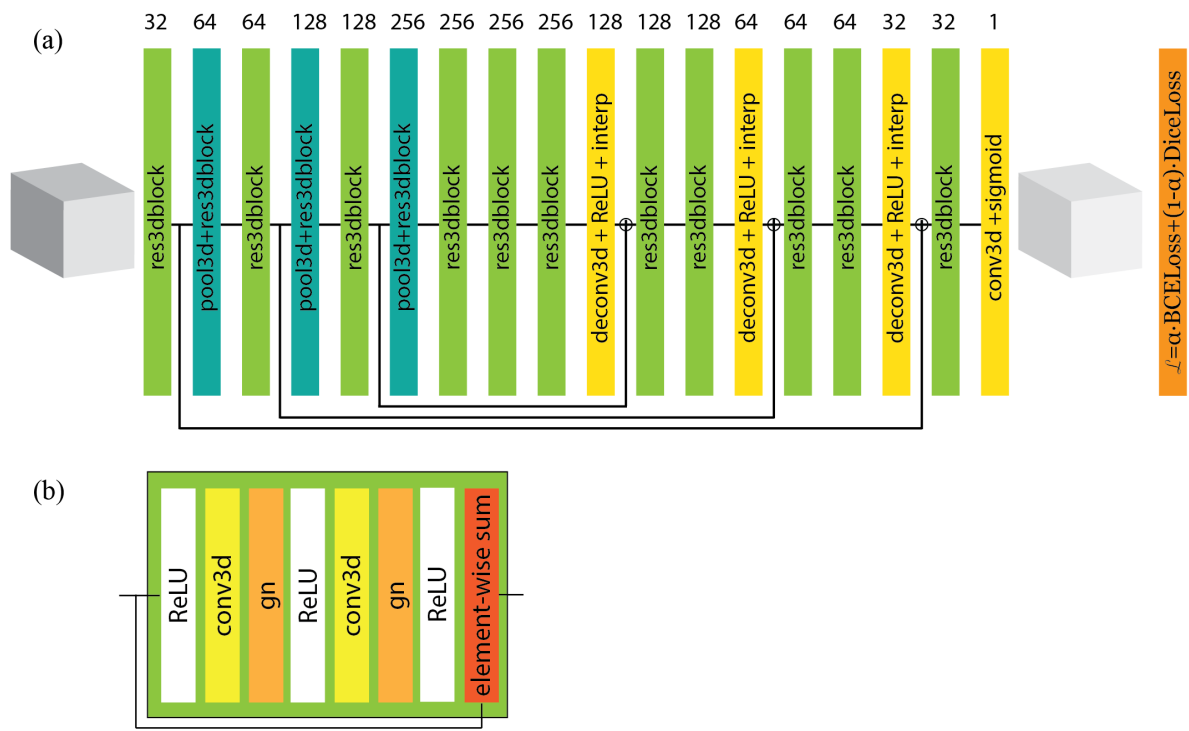

Figure 1. Overview of our 52-layer segmentation model. (a) The network consists of residual blocks (in green), down-sampling blocks (in blue), and up-sampling blocks (in yellow). (2) Our 3D residual block uses a group-normalization (gn) layer to stabilize optimization for a small mini-batch.

**Zoom-In&Out Training Strategy**

To efficiently train our models, we use a two-stage zoom-in&out strategy to first train them on small volumes and then finetune the models on larger volumes. This two-stage training has multiple advantages. First, training with smaller volumes can have a regularizing effect as data augmentation by randomly extracting diverse sub-volumes from original volumes. Second, a "zoom-in" step is an economical option and can utilize sub-optimal graphic processing units (GPUs) for the task. By feeding smaller volumes to older but more accessible GPUs, models can be trained in parallel, and as a result faster. Finally, the "zoom-out" stage involves showing models larger volumes to learn from the broader context of input images that improves the robustness of the model.

**Experimental Settings**

The dataset was split into training, development, and test sets, containing 182; 26; and 31 MRI exams, respectively. In Table 1, we summarized the details and selected hyperparameters for the zoom-in&out stages of the optimization. In evaluation, we cropped out a center $144 \times 172 \times 168$ mm$^3$ volume from the whole MRI scans and fed it to the network to make predictions. All the voxels outside of the cropping window are automatically classified as negative. In addition to our final model at the 150th epoch of finetuning, we built a snapshot ensemble of models at the 50th, 100th, and 150th epoch of zoom-out stage (27). 3D-ResU-Net and 3D-ResU-Net-E denote, respectively, the final model and the snapshot ensemble model. For reproducibility, the complete list of subject IDs in each split is shown in Table E3 in the Appendix.

| Optimization Stage | Zoom-In Stage | Zoom-Out Stage |
|---|---|---|
| **Input volume size (mm$^3$)** | 128×128×128 (24% sub-volume) | 144×172×168 (48% sub-volume) |
| **Training Length** | 1200 epochs | 150 epochs |
| **Initial learning rate** | 1.00E-03 | 1.00E-04 |
| **Optimizer** | Adam optimizer and cosine annealing with warm restart scheduler (28, 29) | |
| **GPU** | Nvidia Titan Xp with 12 GB memory | Nvidia Titan RTX with 24 GB memory |
| **Deep learning framework** | PyTorch (30) | |

Table 1. The details and hyperparameters for the model optimization in our experiments.

**Statistical Analysis**

We evaluate the performance of our segmentation methods on the test set by computing Dice similarity coefficient (DSC), maximal DSC (mDSC ), Hausdorff distance (HD), average symmetric surface distance (ASSD), true positive rate (TPR), and precision for each MRI scan. DSC, HD, and ASSD are computed using a surface distance computation library (31). TPR and precision are computed by using scikit-learn package version 0.21.1 (32), and mDC is implemented according to the algorithm (33). To estimate 95% confidence intervals, we employed bootstrapping for all metrics.

# Results

**Prediction Performance**

Evaluation metrics of our model are summarized in Table 2. 3D-ResU-Net yields an average DSC of 0.64 (0.51-0.74), maximal DSC of 0.66 (0.54-0.76), HD of 20.4 (10.0-33.3) mm, ASSD of 3.6 (1.7-6.2) mm, TPR of 0.81 (0.68-0.90), and precision of 0.62 (0.48-0.74). For 3D-ResU-Net-E, the performance is an average DSC of 0.64 (0.51-0.76), maximal DSC of 0.65 (0.53-0.77), HD of 21.5 (10.0-33.9) mm, ASSD of 3.7 (1.6-6.3) mm, TPR of 0.79 (0.67-0.89), and precision of 0.63 (0.48-0.75). Opposed to previous work (27), the ensemble of our snapshots did not improve the performance. Following these results, we used 3D-ResU-Net for further experiments.

| Methods | DSC | mDSC | HD (mm) | ASSD (mm) | TPR | Precision |
|---|---|---|---|---|---|---|
| 3D-ResU-Net | **0.64 (0.51-0.76)** | **0.66 (0.54-0.76)** | **20.4 (10.0-33.3)** | **3.6 (1.7-6.2)** | **0.81 (0.68-0.90)** | 0.62 (0.48-0.74) |
| 3D-ResU-Net-E | **0.64 (0.51-0.76)** | 0.65 (0.53-0.77) | 21.5 (10.0-33.9) | 3.7 (1.6-6.3) | 0.79 (0.67- 0.89) | **0.63 (0.48-0.75)** |

Table 2. Summary of evaluation metrics. A higher rate is better for DSC, mDSC, TPR, and Precision. For distance metrics (HD and ASSD), a smaller number is better. Best scores are marked in bold.

## Comparison with Existing Methods

We identified recent studies of automatic segmentation that were conducted on the ATLAS dataset and summarized in Table 3. X-Net (16), D-UNet (17), CLCI-Net (15), and our 3D-ResU-Net use specific subsets of the ATLAS data to train and test their model, while Multi-path 2.5D-CNN(18) was trained with two other datasets and tested on the ATLAS dataset. All the models are based on either 2D or 3D U-Net architecture. Among 3D U-Net based models, our 3D-ResU-Net uses significantly larger input volume than that of D-UNet. Although Multi-path 2.5D-CNN takes much larger volume as input than 3D-ResU-Net does, the model does not fully utilize context information in 3D space since the model analyzes input data in 2D and aggregates the results from each slice in the axial plane through post-processing. Due to the large-scale context analysis and our carefully regularized training approach, our method achieved the highest DSC reported using the ATLAS dataset.

| **Methods** | X-Net (16) | Multi-path 2.5D-CNN (18) | D-UNet (17) | CLCI-Net (15) | 3D-ResU-Net (ours) |
|---|---|---|---|---|---|
| **Training data source** | ATLAS | KF & MCW | ATLAS | ATLAS | ATLAS |
| **ATLAS split ratio (train, validation, test) (%)** | 5-fold cross-validation | (0, 0, 100) | (80, 20, 0) | (55, 18, 27) | (76, 11, 13) |
| **Base architecture** | 2D U-Net | 2D U-Net with 3D post-processing | 3D U-Net | 2D U-Net | 3D U-Net |
| **Regularization layers** | Batch normalization | Batch normalization | Batch normalization | Batch normalization | Group normalization |
| **Training strategy** | Adam optimizer, reduce LR on plateau | SGD optimizer, exponential LR decay | SGD optimizer, constant LR | Adam optimizer, constant LR | Adam optimizer, cosine annealing |
| **Loss function** | Dice loss & Cross Entropy | Dice loss | Dice loss & Focal loss | Dice loss | Dice loss & Cross Entropy |
| **Input size (W × H × D)** | 192 × 224 × 1 | 192 × 224 × 192 | 192 × 4 × 192 | 176 × 233 × 1 | 144 × 172 × 168 |
| **Reported DSC** | 0.49 (-) | 0.54 (-) | 0.54 (0.26-0.81) | 0.58 (-) | 0.64 (0.51-0.76) |

Table 3. Summary of different approaches on the ATLAS dataset. evaluation metrics. LR: learning rate. H: height. W: width. D: depth. SGD: stochastic gradient descent. "-" denotes that the corresponding information is not available.

**Qualitative Analysis**

We visualized our automatic segmentation results on the test set by projecting voxel-wise predicted scores onto the original MRI volumes. Figure 2 shows the visualization of reference standard labels and model output viewed from the front-left and front-right side of faces. Visualizations from other samples are also available in Figure E1 in the Appendix. The trained model accurately locates the chronic stroke lesions. Notably, while most of reference standard labels have a chunky structure on the surface, possibly due to the variability of manual human annotations, the predicted lesions tend to have smooth surfaces, which is a realistic assumption for such lesions. We hypothesize that the model has learned this continuous surface from data by internally averaging out marginal voxels of all the training cases and successfully removed variability in human annotation.

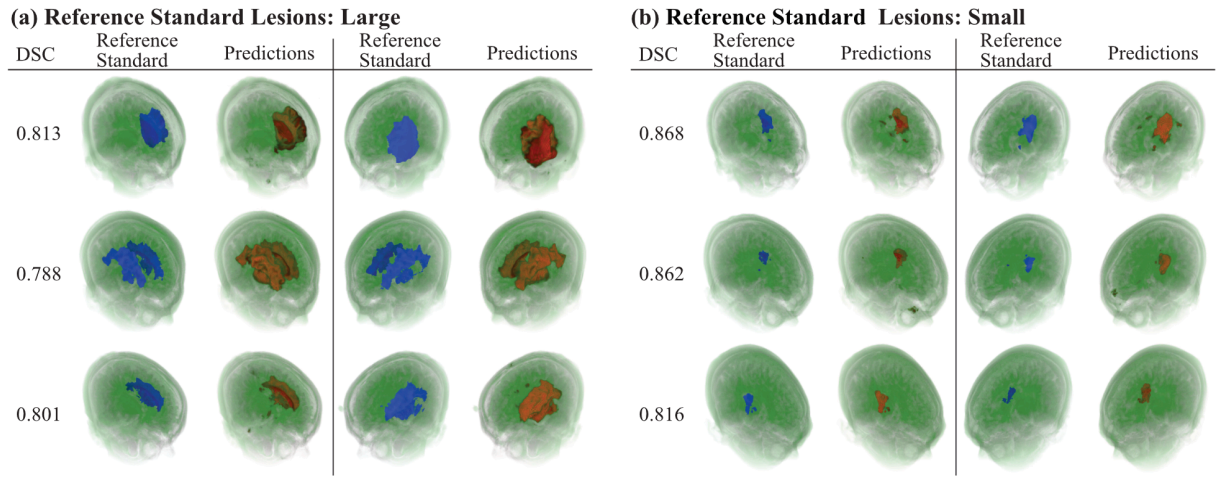

Figure 2. Visualization of reference standard labels (in blue) and lesion predictions by our model (in red). The higher the predicted value is at a voxel, the brighter in red the voxel is. Two groups of samples are shown: large reference labels in (a) and small labels in (b). For each group, the first column is a computed DSC and the rest are visualized reference standards and predictions, from left-front and right-front views. Three typical samples are shown in a row in each group. Best viewed in color.

**Lesion Size and Model Performance**

We further analyzed the performance of the model in relation to the size of target lesions. Figure 3 plots the number of positive voxels in the reference standard and a computed DSC of prediction for each sample in the test set. We observe a trend ($R^2 = 0.34$) in which a sample with large size of tracing has been predicted with a high DSC score. The median of DSC is 0.75, which is 0.11 higher than the average. Table 4 shows a performance summary of the model given a subset of test samples where each subset is composed of a quarter of percentiles when samples are rank-ordered by the number of positive lesion voxels. For example, the first group includes test samples with the size of lesion voxels being smaller than the 25th percentile of the whole test set. The model achieves the highest DSC (0.74-0.84) and TPR (0.79-0.95) on samples with larger positive lesions (75%-100%). In the distance metrics, the model also achieves the lowest mean HD of 13.6 (2.8-35.2) and mean ASSD of 1.8 (0.5-2.8). We confirmed that segmentation performance in both voxel-based and surface-based metrics improves as the lesion to be classified gets larger. The same trend is reported in Ito et al., 2018 and Vorontsov et al., 2019 (34, 35).

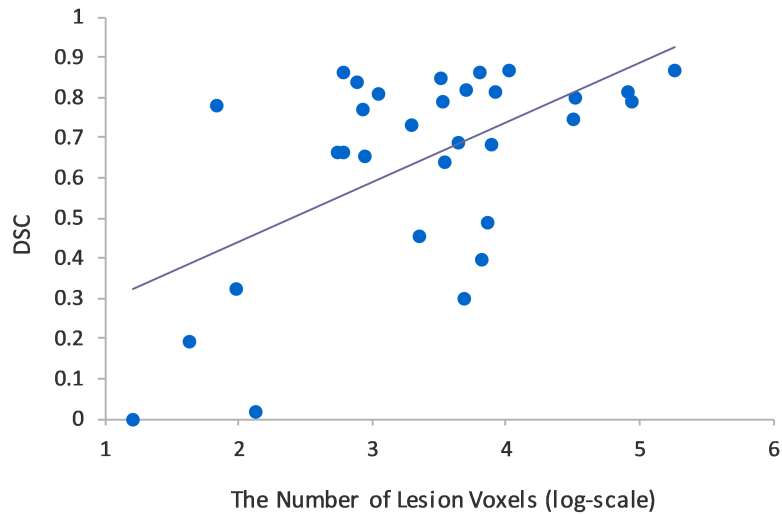

Figure 3. DSC scores and the total number of positive lesion voxels (in base-10 log scale) are computed and plotted for each sample in the test set. $R^2$ of this distribution is 0.34.

| Percentile in per-sample lesion size distribution | DSC | HD (mm) | ASSD (mm) | TPR | Precision |
| --- | --- | --- | --- | --- | --- |
| 0-25% | 0.41 (0.07-0.78) | 23.4 (3.7-51.1) | 4.9 (0.7-13.2) | 0.74 (0.30-0.99) | 0.39 (0.01-0.81) |
| 25-50% | 0.72 (0.59-0.82) | 18.0 (1.8-35.6) | 3.8 (0.5-7.2) | 0.78 (0.64-0.91) | 0.71 (0.53-0.83) |
| 50-75% | 0.62 (0.42-0.80) | 26.1 (2.3-58.0) | 4.1 (0.6-8.8) | 0.82 (0.53-0.96) | 0.60 (0.39-0.79) |
| 75-100% | **0.80 (0.74-0.84)** | **13.6 (2.8-35.2)** | **1.8 (0.8-2.6)** | **0.87 (0.79-0.95)** | **0.75 (0.64-0.84)** |

Table 4. Comparison of evaluation metrics with respect to the subsets of test set samples. Test samples are sorted by the number of positive lesion voxels in increasing order and grouped in four ranges, shown in the first column. DSC, HD, ASSD, TPR, and precision are computed for each group. Best scores are marked in bold.

**Effectiveness of Zoom-in&out Strategy**

To further validate our methodology, we investigate the impact of our zoom-in&out training strategy on the performance of our model by evaluating models with and without finetuning on large volumes. In addition to the metrics we used for our main experiment, we compute the micro-average of DSC (microDSC), which is a global statistic to evaluate the per-voxel performance of our model and is less susceptible to the size of lesions. Here, 3D-ResU-Net-F denotes the model without the finetuning step, distinguished from the 3D-ResU-Net. Table 5 summarizes this ablation study. Through finetuning, the per-voxel and per-sample segmentation performance are improved by 6% and 4%, respectively. The surface distances between the manual tracing and automatic segmentation measured by HD and ASSD are closer by 14.7 mm and 4.0 mm. Except for TPR, the model after the finetuning with larger volumes shows higher performance on all the metrics. Of note, 3D-ResU-Net-F was converged after 1,200 epochs with annealed learning rate, thus we are confident that an additional 150 epochs of training does not improve the performance without increasing the size of input volumes. Also, training models for 1,200 epochs with the zoom-in&out method takes about 5 days, while training entirely with large volumes takes more than 3 weeks. Thus, the zoom-in&out strategy is an effective and viable option for training 3D segmentation models of large 3D T1W MRI images.

| Methods | microDSC | DSC | HD (mm) | ASSD (mm) | TPR | Precision |
|---|---|---|---|---|---|---|
| 3D-ResU-Net-F | 0.73 | 0.60 (0.47-0.73) | 35.1 (20.4-51.3) | 7.6 (3.7-12.3) | **0.83 (0.71-0.91)** | 0.54 (0.39-0.67) |
| 3D-ResU-Net | **0.79** | **0.64 (0.51-0.76)** | **20.4 (10.0-33.3)** | **3.6 (1.7-6.2)** | 0.81 (0.68-0.89) | **0.62 (0.48-0.74)** |
| Δ | +0.06 | +0.04 | -14.7 | -4.0 | -0.02 | +0.08 |

Table 5. Results of our ablation study examining the effect of our zoom-in&out training strategy. Finetuning with larger extracted volumes is applied on a 3D-ResU-Net-F model to obtain a 3D-ResU-Net model. The last row is the difference in performance between the 3D-ResU-Net and 3D-ResU-Net-F for each metric. Best scores are marked in bold.

## Discussion

Identifying lesions and irreversible brain tissue damage on patient MRI scans after a stroke is challenging, especially when the amount of time and resources are limited. In this study, we developed a deep learning model for 3D segmentation of chronic stroke lesions to assist neuroradiologists in this task and further provide personalized rehabilitation for patients to achieve effective recovery. On the test set, the average symmetric surface distance of lesions identified by our segmentation model was 3.6 mm. The average Dice similarity coefficient score of our model was 0.64, with a median of 0.78, which is the highest score achieved on the ATLAS dataset. The overall performance indicates that a 3D deep neural network is a promising method for volumetric segmentation of chronic stroke lesions in T1W MRI.

Our technical contribution in this study is twofold. First, we have established a new state-of-the-art on the ATLAS dataset using the latest deep learning architecture and techniques to further encourage the research in MRI analysis of chronic stroke. Second, we have presented a novel zoom-in&out strategy for effectively training 3D segmentation models in high resolution volumes.

Of note, our study has some limitations. Since the dataset is relatively small, further validations on external datasets of chronic stroke MRI scans are required to verify the generalizability of the segmentation performance. The dataset used in this study contains scans of embolic stroke only, which accounts for the majority of strokes, however, further validation with other types of strokes is worthwhile. In addition, our method experienced the same problem as the previous work in which segmentation performance degraded on volumes with small stroke lesions. Small lesions are reasonably challenging to locate because features of such lesion are subtle and hard to characterize. Notably, missing small lesions of primary stroke would result in a near-zero DSC score because the contribution of each positive voxel is much higher than that of cases with large lesions, and thus leads to having a much lower average DSC than median DSC score. To improve the sensitivity for small lesions, we plan to extend our model to perform at the original resolution for a more detailed feature extraction. The minimum size of lesions in our dataset is 10 $mm^3$, which will be invisible after the first downsampling; thus, to identify such lesions, it is essential to have an extensive feature extraction in the early stage. Of note, by excluding the downsampled path, our model has only five convolutional layers that operate on the original scale. Lastly, our error analysis shown in Figure E2 demonstrates that a few MRI scans in the dataset have visual inconsistencies, possibly introduced at original scanning or image-processing steps in data curation.

Currently, we are considering several avenues for extending our work. From a clinical perspective, the lesion segmentation is a part of the clinical pipeline for providing rehabilitation service for stroke survivors. To fully extend the potential of current research for actionable clinical practice, we plan on building an application that 1) performs the segmentation of chronic stroke lesions, 2) identifies disabled functionalities, and 3) predicts the effectiveness of rehabilitation for each neurological deficits, simultaneously. The first task provides evidence for the second task, and the second task forms a basis for the third task. This pipeline can provide a practical tool that helps clinical decision making. We expect 3D convolutional segmentation architectures are extendable to perform all three tasks. There is evidence that a multi-task model would learn robust features and achieve better performance than models that are trained for a single task (36, 37). To this end, another dataset that records the current status of disabilities, rehabilitation, and recovery of patients would be necessary, in addition to the MRI scans and segmentation masks, for our future work. We expect that our study will establish a standard in this domain and promotes further research to advance current state-of-the-art methodologies for volumetric segmentation of chronic ischemic stroke lesion on T1W MRI scans.

## Acknowledgments

This research was supported in part by National Institute of Health grant (R01LM012837). Data for this work were obtained from ICPSR website,[1] (Principal Investigator: Sook-Lei Liew), supported by NIH-funded Center for Large Data Research and Data Sharing in Rehabilitation (P2CHD06570 and NIH 1K01HD091283). The authors would like to thank Lamar Moss, Maksim Bolonkin, Jason Wei, and Jerry Wei for their feedback on the manuscript.

---

[1] https://www.icpsr.umich.edu/icpsrweb/ADDEP/studies/36684